\documentclass[12pt,fleqn]{article}

\usepackage{a4wide,amsmath,axodraw}

\begin{document}

\newcommand{\nl}{\nonumber \\}
\newcommand{\suml}{\sum\limits}
\newcommand{\eqn}[1]{Eq.~\!(\ref{#1})}
\newcommand{\fig}[1]{Fig.~\!\ref{#1}.}
\newcommand{\tab}[1]{Tab.~\!\ref{#1}}
\newcommand{\ep}{\epsilon}
\newcommand{\al}{\alpha}
\newcommand{\alasym}{\alpha^{\textrm{asym}}}
\newcommand{\sss}[1]{\scriptscriptstyle{#1}}

\title{{\bf Computer-aided analysis of\\ Riemann sheet structures}}

\author{
Andr\'{e} van Hameren\thanks{andrevh@sci.kun.nl}~ and 
Ronald Kleiss\thanks{kleiss@sci.kun.nl}\\
University of Nijmegen, Nijmegen, the Netherlands}

\maketitle

\begin{abstract}
We report on experience with an investigation of the analytic structure
of the solution of certain algebraic complex equations. In particular
the behavior of their series expansions around the origin is discussed.
The investigation imposes the need for an analysis of the singularities and 
the Riemann sheets of the solution, in which numerical methods are used.

\end{abstract}

\thispagestyle{empty}

\newpage
\pagestyle{plain}
\setcounter{page}{1}

\section{Introduction}
In various problems in theoretical physics we have encountered problems
that are formulated as follows: consider an entire function $F$ of
a complex number $y$, such that
\begin{equation}
F(y) \sim y^m \quad\textrm{as $y\to0$}\;\;,
\end{equation}
with nonnegative integer $m$ (in practice, we have met cases with
$m=1$ and $m=2$). The task at hand is then to find 
information about $y=y(x)$ such
that 
\begin{equation}
F(y(x)) = x^m\;\;.
\label{eqnone}
\end{equation}
In general, both the form of the series expansion of $y(x)$ around
$x=0$ and the nature of its singularities are of interest.
Such questions arise, for instance, in the combinatorial
problem of determining the number of Feynman diagrams contributing to given 
scattering amplitudes in various quantum field theories \cite{counting}, 
in the computation of the oscillation period of 
nontrivial one-dimensional potential wells \cite{discrep}, 
in the statistical bootstrap model for hot hadronic matter 
(refs. in \cite{hagedorn} and for a recent discussion see \cite{kapoyannis}), 
and in renormalization theory connected with the 't Hooft transformation 
\cite{khuri}.
An important and interesting example, studied in 
detail in \cite{hagedorn}, is the so-called {\em bootstrap equation}:
\begin{equation}
F_B(y) = 2y + 1 - e^y\;\;,
\label{bootstrap}
\end{equation}
which obviously has $m=1$.
We shall consider functions $F$ of the more general form
\begin{equation}
F(y) = P(y) + e^yQ(y)\;\;,
\end{equation}
where $P$ and $Q$ are polynomials of finite degree $d_P >0$ 
and $d_Q \ge0$, respectively, with
real coefficients. As our working example, taken from \cite{discrep},
we shall consider the function $F_1$ defined as
\begin{equation}
F_1(y) = - 2 - 2y + 2e^y\;\;,
\label{ours}
\end{equation}
for which $m=2$.
In this paper we shall concentrate on the analysis 
of the Riemann sheet structure of 
those solutions of these equations that have a series
expansion around $x=0$. 
To determine the asymptotic behavior of these expansions,
the nature of the singularities will be analyzed numerically.
The results are justified by the fact that, in our calculations, only
finite computer accuracy is required, as we shall demonstrate.

\section{Identification of the Riemann sheets}
As a first step we identify the various Riemann sheets by their value of
$y(0)$: the sheet labeled $s$ will have $y(0)=Y_s$ for that sheet.
Obviously, $y(0)=0$ is a solution with multiplicity $m$. In general,
there will be $d_P$ solutions if $Q(y)=0$, and infinitely many if $Q$
is non-vanishing. It will be helpful if we can identify the Riemann sheet
on which pairs $(x,y(x))$ lie when $x$ is small but nonzero. 
This is indeed possible, and
we shall illustrate it using $F_1$. Let us write $y=\xi + i\eta$ with 
$\xi$ and $\eta$ real numbers. We are then looking for solutions of
$F_1(\xi+i\eta)=0$, or
\begin{equation}
\xi = \log\left(\frac{\eta}{\sin\eta}\right)\;\;\;,\;\;\;
1 + \log\left(\frac{\eta}{\sin\eta}\right) - \frac{\eta}{\tan\eta} = 0\;\;.
\end{equation}
Inspecting the left-hand side of the last equation, we can immediately
see that its zeroes are quite nicely distributed. We can usefully
enumerate them as Im$(Y_s) = u_s$, where  the sheet number $s$
takes only the odd integer values $\pm1,\pm3,\pm5,\ldots$.
For positive $s$, the zero $u_s$ is certainly located in the interval
where $\sin u_s>0$, {\it i.e.\/} $(s-1)\pi\le u_s<s\pi$,
and $u_{-s}=-u_s$. We have $u_1=u_{-1}=0$, and
for increasing $s$ the zero $u_s$ moves upwards in its interval, until
asymptotically we have 
$u_s \sim a_s - (\log a_s)/a_s$ with $a_s = (s-1/2)\pi$.
In \tab{table1} we give the values of $Y_s$ for $F_1$, for the first
few values of $s$.
\begin{table}
\begin{center}
\begin{tabular}[b]{|l|c|}
\hline
$s$ & $Y_s/\pi$ \\ \hline
    1 &  (    0.0000,    0.0000 )\\
    3 &  (    0.6649,    2.3751 )\\
    5 &  (    0.8480,    4.4178 )\\
    7 &  (    0.9633,    6.4374 )\\
    9 &  (    1.0478,    8.4490 )\\
   11&   (    1.1145,   10.4567 )\\
\hline
\end{tabular}
\caption[.]{The first few Riemann sheet solutions for $F_1(Y_s)=0$.}
\label{table1}
\end{center}
\end{table}
Because the values $Y_s$ fall in disjoint intervals, 
for small $x$ we need to know $y(x)$ only to a limited
accuracy in order to be able to identify its Riemann sheet. The only
nontrivial case is that of sheets $-1$ and $1$, where it is sufficient to
consider the complex arguments:
for $\arg(x)-\arg(y) = 0$ we are on sheet $1$, 
for $|\arg(x)-\arg(y)|=\pi$ we are on sheet $-1$.
Again, limited computer accuracy is acceptable here, and for larger $m$
we simply have $m$ different values of the argument, distinguished in
an analogous manner. 
Note that of course the
labeling of the sheets is rather arbitrary: we have chosen the odd integers
in order to emphasize that both sheet $1$ and $-1$ can be considered the
principal Riemann sheet. For the bootstrap equation (\ref{bootstrap})
it is more natural to label the single principal Riemann sheet with
$y(0)=0$ as sheet number zero.

\section{Series expansion}
We want to compute $y(x)$ as a Taylor series around $x=0$:
\begin{equation}
y(x) = \suml_{n\ge0}\alpha_nx^n\;\;.
\label{expan}
\end{equation}
Obviously, $\alpha_0$ can be chosen as one of the $u_s$ above. On principal
sheets, with $\alpha_0=0$, we also have immediately that $\alpha_1$ must be
chosen out of the $m$ possibilities with $\alpha_1^m=1$. The other
coefficients must then be computed (algebraically or numerically) by some
recursive method, which we shall now discuss.

It would be straightforward to plug the expansion (\ref{expan}) into
\eqn{eqnone} and equate the powers of $x$ on both sides, but notice that,
for $Q$ non-vanishing, the number of possible products of coefficients
grows very rapidly, so that the computer time
needed to find the first $N$ coefficients grows exponentially
with $N$. As already mentioned in \cite{hagedorn}, 
the better way is to differentiate
\eqn{eqnone} with respect to $x$ so that we obtain the
nonlinear differential equation
\begin{equation}
y'(x)\left[P'(y)Q(y) + (Q(y)+Q'(y))(x^m-P(y))\right] = mx^{m-1}Q(y)\;\;.
\end{equation}
This equation yields a recursion relation involving 
products of at most $d_P+d_Q+1$ coefficients, 
so that a truncated power series can be computed in polynomial time.
As an example, for $F_1$ we find the following differential equation:
\begin{equation}
y'(x)(x^2+ 2y(x)) = 2x\;\;,
\end{equation}
and the following recursion relation:
\begin{align}
&\alpha_0\alpha_1 =  0\;\;\;,\;\;\;
2\alpha_0\alpha_2 + \alpha_1^2 - 1  =  0\;\;,\nl
&n\alpha_0\alpha_n + (n-2)\alpha_{n-2} + 
2\suml_{p=1}^{n-1}p\alpha_p\alpha_{n-p}  =  0\;\;,\;\;n\ge3\;\;.
\label{ourrec}
\end{align}
We see immediately that $y(x)$ is necessarily even in $x$ if
$\alpha_0\ne0$, {\it i.e.\/} on the non-principal Riemann sheets.
In that case, we also see that if $\al_n$, $n=0,2,\ldots$ is a solution, 
then also $\al^*_n$, $n=0,2,\ldots$ is a solution, where the asterix 
stands for complex conjugation.  This is a result of the fact that 
if $y(x)$ is a solution of \eqn{ours}, then also $y^*(x^*)$ is a solution. 
In practice, these solutions 
give the function values on the different Riemann sheets of one solution. The 
analysis of the previous section proves that $y_s(0)=y_{-s}(0)^*$ so that 
the solutions satisfy $y_s^*(x)=y_{-s}(x^*)$ and the expansion coefficients 
satisfy
\begin{equation}
   \al^{(s)}_n=(\al^{(-s)}_n)^* \;\;.
\label{coefstar}   
\end{equation}
On the principal Riemann sheets we have 
$\alpha_0=0$ and $\alpha_1^2=1$ as mentioned,
and the two solutions on sheet $1$ and sheet $-1$ are related by
$y_{-1}(x) = y_1(-x)$. 
For $y_1(x)$ we find, finally:
\begin{equation}
\alpha_n = -\frac{1}{2(n+1)}
\left[(n-1)\alpha_{n-1} +
 2\suml_{p=2}^{n-1}p\alpha_p\alpha_{n+1-p}\right]\;\;,
\end{equation}
for $n\ge2$. Using this relation we have been able to compute many thousands
of terms. The recursion appears
to be stable in the forward direction, but we
have not tried to prove this or examine the stability in the general case.

In series expansions it is of course always important to know the
convergence properties or, equivalently, the asymptotic behavior
of $\alpha_n$ as $n$ becomes very large. In the next section, we therefore
turn to the singularity structure of $y(x)$.

\section{Singularities and branches}
In order to find information about the singularity structure of $y(x)$, we
employ the techniques developed in \cite{counting}, which we recapitulate here.
Singularities are situated at those values $y_k$ of $y$ where
\begin{equation}
F'(y_k) = 0\;\;.
\end{equation}
Since $F$ is entire we also know that these singular points must form an
enumerable set, {\it i.e.} we can find, and label, them as distinct points.
We shall assume that these singularities are square-root branch points,
for which it is necessary that
\begin{equation}
F''(y_k) \ne 0\;\;,
\end{equation}
If $F''$ vanishes at $y_k$ but $F'''$ does not, 
we have a cube-root branch point,
and so on. If, for non-vanishing $Q$, 
all derivatives vanish (as for instance when $F(y)=e^y$) we
have, of course, a logarithmic branch point. We know that $y=-\infty$ 
corresponds to a logarithmic branch point, and it is 
to remove this to infinity
in the $x$ plane that we have required $d_P>0$. In our
examples all the singularities at finite $x$ will be square-root
branch points. 
The position of the singularity in the $x$ plane, $x_k$, is of course
given by
\begin{equation}
F(y_k) = x_k^m\;\;,
\end{equation}
so that there are $m$ different possible positions, lying
equally spaced on a circle around the origin. 
We shall denote them by $x_{k,p}$ with $p=1,2,\ldots,m$.
Note that, in first instance, it is not clear at all whether $x_{k,p}$ for 
certain $k$ and $p$ is indeed a singular point on a specific Riemann sheet. 
Later on, we shall describe how to determine this numerically.
For values of $x$ close to an observed singular point
$x_{k,p}$ we may expand the left-hand and 
right-hand side of \eqn{eqnone} to obtain
\begin{equation}
\frac{1}{2}(y-y_k)^2F''(y_k) \sim mF(y_k)\left(\frac{x}{ x_{k,p}}-1\right)\;\;,
\label{approx}
\end{equation}
where we have dropped the higher derivative terms. Very close to the
branch point we may therefore approximate $y(x)$ by
\begin{equation}
y(x)  \sim  y_k + \beta_{k,p}\;\left(1-\frac{x}{ x_{k,p}}\right)^{1/2}
\;\;,\quad
\beta_{k,p}^2  \equiv  -\frac{2mF(y_k)}{ F''(y_k)}\;\;.
\end{equation}
Note that there are only two possible values for $\beta_{k,p}$,
and each singular point $x_{k,p}$ goes with one or the other of these.
Again numerical methods will help in determining which one of the two is 
the correct choice.

We are now in a position to compute the asymptotic behavior of the
coefficients $\alpha_n$. To find it, we first determine, {\em for a
given Riemann sheet}, which are the $x_{k,p}$ that lie closest to the
origin: this gives us the radius of convergence of the expansion
of $y(x)$ in that Riemann sheet. We then have to determine those $p$
for which $x_{k,p}$ is actually a singular point. We shall do this
numerically, in the way described in the following section. Let us denote
the set of values of $p$ for which this is the case by ${\cal P}$.
Now, we may use the fact that
\begin{align}
&\sqrt{1-x}  =  1 - \suml_{n\ge1}\gamma_n x^n\;\;\;,\;\;\;
\gamma_n = \frac{(2n-2)!}{2^{2n-1}(n-1)!n!}\;\;,\nl
&\gamma_n    \sim \frac{1}{\sqrt{4\pi}}\,n^{-3/2} + {\cal O}(n^{-5/2})\;\;,\;\;
n\to\infty\;\;,
\end{align}
where we have chosen that square root that is real and 
positive for $1-x$ real and 
positive. The asymptotic behavior of $\alpha_n$ as $n\to\infty$ 
must therefore be given by
\begin{equation}
\alpha_n \sim \frac{-1}{ n^{3/2}\sqrt{4\pi}}
\suml_{p\in{\cal P}} \frac{\beta_{k,p}}{ x_{k,p}^n}\;\;.
\label{asymptot}
\end{equation}
Amongst other things, this provides a powerful numerical check on the
accuracy of the $\alpha_n$ as computed by the recursive technique.
We shall now discuss how the singularity structure of our problem can be
investigated numerically.

\section{Computer searches for sheet structures}
The main tool we use for our computer studies is a method for
taking small steps over a Riemann sheet, that is, given the fact that
for some value $x_1$ the point
$y_1=y(x_1)$ is determined to belong to a certain Riemann sheet, we perform
a small step $\Delta x$ to a point $x_2$
and find the point $y_2 = y(x_2)$
on the same Riemann sheet. Our method to do this is nothing but
Newton-Raphson iteration: we simply iterate the mapping
\begin{equation}
y \leftarrow y - \frac{F(y) - x_2^m}{ F'(y)}\;\;,
\end{equation}
until satisfactory convergence is obtained. The starting value
for this iteration is just the point $y_1$. A few remarks are in
order here. In the first place, it must be noted that for this method to
work, $y_1$ must be in the basin of attraction of $y_2$. Since, except
at the branch points, which we shall expressly avoid, $y(x)$ is
a continuous and differentiable function of $x$, this can always be
arranged by taking $\Delta x$ small enough. In the second place, the
accuracy with which $y_1$ is actually a solution of \eqn{eqnone} is
not important as long as it is in the basin of attraction of $y_2$: 
therefore, there is no buildup of numerical errors in this method
if we restrict ourselves to just keeping track of which Riemann sheet
we are on. Finally, problems could arise if two Riemann sheet values
of $y$ for the same $x$ are very close. But, since $F$ is
an entire function, we know that the solutions of \eqn{eqnone}
must either completely coincide or be separated by a finite
distance, any inadvertent jump from one sheet to another can be detected
and cured by, again, taking a small enough $\Delta x$.

We have applied the following method for detecting and characterizing the
various singular points. We start on a Riemann sheet $s_1$ at a value
$x$ close to zero, and determine $y(x)$ on that Riemann sheet. We then let
the parameter $x$ follow a prescribed contour that circles a selected
would-be singularity $x_{k,p}$ once (and no other singularities), and then
returns to the starting point close to the origin. We then determine
to which Riemann sheet the resulting $y$ belongs. In this way we can find 
whether $x_{k,p}$ is, in fact, a singular point for the starting sheet,
and, if so, which two sheets are connected there. It is also possible, of
course, to certify the square-root branch point nature of a singular point
by circling twice around it, and checking that one returns to the
original Riemann sheet.

One important remark is in order here. In our tracking over the Riemann
sheet, it is necessary that we do not cross branch cuts (except of course
the one connected to the putative singularity). Since these branch cuts
can be moved around in the complex $x$ plane, {\em the contour
chosen defines the (relative) position of the branch cuts}. 
The sheets that are said to be connected at a particular branch cut
are therefore also determined by the choice of contour. 
Of course, choosing a different contour will change the
whole system of interconnected sheets in a consistent
manner, so that in fact, given one choice of contour and its system
of sheets, we can work out what system of sheets will
correspond to another choice of contour.
We shall illustrate this in the following.

Suppose, now, that $x_{k,p}$ is one of the singular points on a certain
sheet that is closest to the origin. We can then follow, on that sheet,
a straight line running from $x_1$ close to the origin to a point 
$x_2$ for which $x_2/x_{k,p}$ is real and just a bit smaller than one.
Since $x_{k,p}$ is by assumption closest to the origin, there is then
no ambiguity involved in determining which one of the two possible
complex arguments of $\beta_{k,p}$ we have to take. Thus, we can find
all the information needed to compute the asymptotic behavior of
$\alpha_n$ on that sheet.

\section{An example}
Having established the necessary machinery, we shall now discuss a concrete
example of our method. For this, we have taken the function $F_1$
of \eqn{ours}. It is, in fact, closely related to the 
very well-understood bootstrap
equation (\ref{bootstrap}): by substituting, in \eqn{bootstrap},
$y\to \log2+y$ and $x\to2\log2 - 1 -x^2$, we obtain \eqn{ours}.
Its Riemann sheet structure, however, is quite different, as we shall see.
Note that under this transformation, the origin $x=0$, $y=0$ for $F_1$ 
corresponds to the first singularity in $F_B$.

\subsection{The singularities}
The values of $y(0)$ on the different Riemann sheets for $F_1$, 
namely $Y_s$ for $s=\pm1,\pm3,\ldots$ have already
been discussed above. The singular values $y_k$ are simply given
by
\begin{equation}
F_1'(y_k) = 2e^{y_k} - 2 =0\;\;\;\Rightarrow\;\;\; y_k = 2i\pi k\;\;,
\end{equation}
so that the possible singular points $x_{k,p}$ satisfy
\begin{equation}
x_{k,p}^2 = -4i\pi k\;\;.
\end{equation}
Note that $k=0$ does not correspond to a singular point. 
The positions of the possible singularities in the complex $x$ plane
are therefore as follows. For positive integer $k$:
\begin{align}
&x_{k,1}  =  iz_k\;\;\;,\;\;\;x_{k,2} = -iz_k\;\;\;,\nl
&x_{-k,1}  =  z_k\;\;\;,\;\;\;x_{-k,2} = -z_k\;\;\;,\nl
&z_k  =  (1+i)\sqrt{2\pi k}\;\;.
\end{align}
At all these various possible singularities, we have
\begin{equation}
\beta_{k,p}^2 = 8i\pi k\;\;,
\end{equation}
and therefore we may write
\begin{align}
&\textrm{for $k>0$}:\quad \beta_{k,p} = \ep_{k,p}(1+i)\sqrt{4\pi|k|}\;\;,\nl
&\textrm{for $k<0$}:\quad \beta_{k,p} = \ep_{k,p}(1-i)\sqrt{4\pi|k|}\;\;,
\end{align}
where the only number to be determined is $\ep_{k,p}\in\{-1,1\}$. It must be 
kept in mind
that the value of $\ep$ depends of course on the sheet: we take the
convention that we work on the sheet with the lowest number
(in absolute value). When viewed from the other sheet, the value of
$\ep$ is simply opposite.

\subsection{The Riemann sheet structure}
We now have to discuss how the branch cuts should run in the complex $x$
plane. There are two simple options (and an infinity of more  complicated
ones): in the first option (I), we choose to let the branch cuts
extend away from the origin parallel to the real axis. This corresponds
to tracking a contour that, say, first moves in the imaginary direction,
and then in the real direction, to arrive close to the chosen singularity.
The other option (II) is to take the cuts parallel to the imaginary
axis, so that a contour that does not cross branch cuts {\it en route\/}
first goes in the real direction, and then in the imaginary direction.
Note that these two alternatives do, indeed, correspond to different
implied relative positionings of the branch cuts.
\begin{figure}
\begin{center} 
\begin{picture}(120.0,120.0)(35,0)
\LinAxis(  0.0,  0.0)(120.0,  0.0)(1.0,1,3,0,1)
\LinAxis(  0.0,  0.0)(  0.0,120.0)(1.0,1,-3,0,1)
\LinAxis(  0.0,120.0)(120.0,120.0)(1.0,1,-3,0,1)
\LinAxis(120.0,  0.0)(120.0,120.0)(1.0,1,3,0,1)
\Text(  0.0,-10)[]{$-5$} \Text( 60.0,-10)[]{$0$} \Text(120.0,-10)[]{$5$}
\Text(-10,  0.0)[]{$-5$} \Text(-10, 60.0)[]{$0$} \Text(-10,120.0)[]{$5$}
\Vertex( 90.08, 90.08){1} \Vertex(102.54,102.54){1} \Vertex(112.10,112.10){1}
\Vertex( 29.92, 90.08){1} \Vertex( 17.46,102.54){1} \Vertex(  7.90,112.10){1}
\Vertex( 29.92, 29.92){1} \Vertex( 17.46, 17.46){1} \Vertex(  7.90,  7.90){1}
\Vertex( 90.08, 29.92){1} \Vertex(102.54, 17.46){1} \Vertex(112.10,  7.90){1}
\Text( 86.08, 89.08)[r]{$\sss{(-1,1)}$} \Text( 98.54,101.54)[r]{$\sss{(-2,1)}$}
\Text(108.10,111.10)[r]{$\sss{(-3,1)}$} \Text( 33.92, 89.08)[l]{$\sss{(1,1)}$}
\Text( 21.46,101.54)[l]{$\sss{(2,1)}$} \Text( 11.90,111.10)[l]{$\sss{(3,1)}$}
\Text( 33.92, 30.92)[l]{$\sss{(-1,2)}$} \Text( 21.46, 18.46)[l]{$\sss{(-2,2)}$}
\Text( 11.90,  8.90)[l]{$\sss{(-3,2)}$} \Text( 86.08, 30.92)[r]{$\sss{(1,2)}$}
\Text( 98.54, 18.46)[r]{$\sss{(2,2)}$} \Text(108.10,  8.90)[r]{$\sss{(3,2)}$}
\Text(60,-30)[]{a: The numbering $(k,p)$ of the singularities.}
\DashLine(60,0)(60,120){1.5}
\DashLine(0,60)(120,60){1.5}
\end{picture}     
\begin{picture}(120.0,120.0)(-60,0)
\LinAxis(  0.0,  0.0)(120.0,  0.0)(1.0,1,3,0,1)
\LinAxis(  0.0,  0.0)(  0.0,120.0)(1.0,1,-3,0,1)
\LinAxis(  0.0,120.0)(120.0,120.0)(1.0,1,-3,0,1)
\LinAxis(120.0,  0.0)(120.0,120.0)(1.0,1,3,0,1)
\Text(  0.0,-10)[]{$-5$} \Text(100.0,-10)[]{$0$} \Text(120.0,-10)[]{$1$}
\Text(-10,  0.0)[]{$-1$} \Text(-10, 20.0)[]{$0$} \Text(-10,120.0)[]{$5$}
\Text(60,-30)[]{b: $x$-plane}
\DashLine(100,0)(100,120){1.5} \DashLine(0,20)(120,20){1.5}
\Line( 99.00, 21.00)( 99.00, 90.90) \Line( 99.00, 90.90)( 39.10, 90.90)
\CArc( 29.10, 90.90)(10,0,360)
\Vertex( 49.87, 70.13){1}\DashLine(0, 70.13)( 49.87, 70.13){6}
\Vertex( 29.10, 90.90){1}\DashLine(0, 90.90)( 29.10, 90.90){6}
\Vertex( 13.17,106.83){1}\DashLine(0,106.83)( 13.17,106.83){6}
\end{picture}   \\
\begin{picture}(120.0,120.0)(25,60)  
\LinAxis(  0.0,  0.0)(120.0,  0.0)(1.0,1,3,0,1)
\LinAxis(  0.0,  0.0)(  0.0,120.0)(1.0,1,-3,0,1)
\LinAxis(  0.0,120.0)(120.0,120.0)(1.0,1,-3,0,1)
\LinAxis(120.0,  0.0)(120.0,120.0)(1.0,1,3,0,1)
\Text(  0.0,-10)[]{$-2$} \Text( 34.3,-10)[]{$0$} \Text(120.0,-10)[]{$5$}
\Text(-10,  0.0)[]{$-1$} \Text(-10, 17.1)[]{$0$} \Text(-10,120.0)[]{$6$}
\Text(60,-30)[]{c: sheet 1, single loop}
\Curve{( 33.43, 18.01) ( 33.45, 18.99) ( 33.49, 19.90) ( 33.55, 20.81)
       ( 33.62, 21.72) ( 33.82, 23.53) ( 34.07, 25.33) ( 34.39, 27.12) 
       ( 34.77, 28.89) ( 35.21, 30.63) ( 35.71, 32.35) ( 36.27, 34.04) 
       ( 36.88, 35.70) ( 37.55, 37.33) ( 38.27, 38.92) ( 39.03, 40.47) 
       ( 39.87, 42.03) ( 40.79, 43.59) ( 41.75, 45.09) ( 42.75, 46.54) 
       ( 43.79, 47.93) ( 44.90, 49.30) ( 46.08, 50.65) ( 47.28, 51.93) 
       ( 48.53, 53.18) ( 49.84, 54.38) ( 51.17, 55.51) ( 52.57, 56.62) 
       ( 53.98, 57.65) ( 55.45, 58.65) ( 56.91, 59.58) ( 57.67, 60.03) 
       ( 58.43, 60.46) ( 59.21, 60.90) ( 59.99, 61.31)}
\Curve{( 55.15, 79.81) ( 55.17, 78.93) ( 55.20, 78.07) ( 55.26, 77.17)
       ( 55.34, 76.29) ( 55.56, 74.56) ( 55.87, 72.80) ( 56.25, 71.11) 
       ( 56.71, 69.40) ( 57.24, 67.72) ( 57.84, 66.08) ( 58.50, 64.45) 
       ( 58.85, 63.66) ( 59.22, 62.86) ( 59.60, 62.07) ( 59.99, 61.31)}
\Curve{( 55.15, 79.81) ( 55.16, 80.69) ( 55.20, 81.58) ( 55.26, 82.48)
       ( 55.34, 83.33) ( 55.45, 84.19) ( 55.58, 85.06) ( 55.74, 85.92)
       ( 55.93, 86.78) ( 56.15, 87.64) ( 56.39, 88.50) ( 56.67, 89.35)
       ( 56.97, 90.19) ( 57.30, 91.02) ( 57.66, 91.84) ( 58.04, 92.64)
       ( 58.45, 93.43) ( 58.89, 94.19) ( 59.35, 94.94) ( 59.83, 95.66)
       ( 60.36, 96.39) ( 60.90, 97.10) ( 61.46, 97.77) ( 62.03, 98.42)
       ( 62.65, 99.07) ( 63.27, 99.69) ( 63.93,100.30) ( 64.59,100.88)
       ( 65.03,101.18) ( 65.64,100.55) ( 66.36,100.01) ( 67.12, 99.60)
       ( 67.94, 99.31) ( 68.80, 99.15) ( 69.68, 99.12) ( 70.56, 99.22)
       ( 71.42, 99.46) ( 72.20, 99.82) ( 72.93,100.30) ( 73.58,100.89)
       ( 74.11,101.55) ( 74.55,102.31) ( 74.85,103.11) ( 75.03,103.97)
       ( 75.07,104.85)}
\Curve{( 63.73,104.64) ( 63.79,105.52) ( 63.98,106.38) ( 64.31,107.19)
       ( 64.75,107.93) ( 65.32,108.61) ( 65.97,109.17) ( 66.72,109.63)
       ( 67.53,109.98) ( 68.39,110.20) ( 69.28,110.28) ( 70.16,110.23)
       ( 71.03,110.04) ( 71.86,109.72) ( 72.63,109.27) ( 73.30,108.73)
       ( 73.89,108.09) ( 74.38,107.36) ( 74.74,106.57) ( 74.98,105.72)
       ( 75.07,104.85)}
\Curve{( 63.73,104.64) ( 63.82,103.77) ( 64.04,102.93) ( 64.40,102.12)
       ( 64.87,101.38)}
\DashLine(34.3,0)(34.3,120){1.5} \DashLine(0, 17.1)(120, 17.1){1.5}       
\end{picture}   
\begin{picture}(120.0,120.0)(-5,60)
\LinAxis(  0.0,  0.0)(120.0,  0.0)(1.0,1,3,0,1)
\LinAxis(  0.0,  0.0)(  0.0,120.0)(1.0,1,-3,0,1)
\LinAxis(  0.0,120.0)(120.0,120.0)(1.0,1,-3,0,1)
\LinAxis(120.0,  0.0)(120.0,120.0)(1.0,1,3,0,1)
\Text(  0.0,-10)[]{$ -4$} \Text( 53.3,-10)[]{$  0$} \Text(120.0,-10)[]{$  5$}
\Text(-10,  0.0)[]{$  7$} \Text(-10,120.0)[]{$ 16$}
\Text(60,-30)[]{d: sheet 3, single loop}
\DashLine( 53.3,0)( 53.3,120){1.5}
\Curve{( 80.53, 10.09) ( 80.58,  8.75) ( 80.79,  7.43) ( 81.18,  6.15)}
\Curve{( 80.53, 10.09) ( 80.61, 11.43) ( 80.82, 12.76) ( 81.16, 14.05)
       ( 81.62, 15.32)(82.43,16.2)}
\Curve{( 52.70, 42.16) ( 52.74, 41.74) ( 53.26, 40.45) ( 53.84, 39.14)
       ( 54.43, 37.91) ( 55.09, 36.67) ( 55.76, 35.50) ( 56.49, 34.33)
       ( 57.27, 33.17) ( 58.07, 32.09) ( 58.91, 31.03) ( 59.79, 30.00)
       ( 60.72, 29.00) ( 61.68, 28.04) ( 62.67, 27.12) ( 63.69, 26.25)
       ( 64.76, 25.41) ( 65.87, 24.59) ( 66.98, 23.83) ( 68.11, 23.11)
       ( 69.29, 22.42) ( 70.47, 21.76) ( 71.66, 21.14) ( 72.86, 20.55)
       ( 74.07, 19.98) ( 75.29, 19.42) ( 76.52, 18.87) ( 77.75, 18.33)
       ( 78.98, 17.78) ( 80.21, 17.22) ( 81.43, 16.64) (82.43,16.2)}
\Curve{( 52.70, 42.16) ( 53.98, 42.67) ( 55.29, 43.24) ( 56.53, 43.85)
       ( 57.71, 44.47) ( 58.91, 45.17) ( 60.04, 45.89) ( 61.18, 46.67)
       ( 62.26, 47.47) ( 63.34, 48.34) ( 64.36, 49.21) ( 65.37, 50.15)
       ( 66.31, 51.11) ( 67.24, 52.12) ( 68.11, 53.14) ( 68.96, 54.23)
       ( 69.74, 55.32) ( 70.50, 56.47) ( 71.19, 57.62) ( 71.85, 58.83)
       ( 72.44, 60.04) ( 72.99, 61.30) ( 73.48, 62.57) ( 73.91, 63.88)
       ( 74.29, 65.18) ( 74.59, 66.48) ( 74.84, 67.83) ( 75.03, 69.16)
       ( 75.16, 70.54) ( 75.23, 71.90) ( 75.24, 73.24)}
\Curve{( 68.99, 93.02) ( 69.26, 92.48) ( 69.94, 91.31) ( 70.60, 90.11)
       ( 71.22, 88.88) ( 71.79, 87.66) ( 72.33, 86.41) ( 72.83, 85.16)
       ( 73.29, 83.87) ( 73.71, 82.58) ( 74.08, 81.28) ( 74.40, 79.98)
       ( 74.67, 78.66) ( 74.90, 77.33) ( 75.07, 75.98) ( 75.18, 74.62)
       ( 75.24, 73.24)}
\Curve{( 68.99, 93.02) ( 70.16, 93.72) ( 71.36, 94.39) ( 72.57, 95.00)
       ( 73.79, 95.55) ( 75.03, 96.06) ( 76.30, 96.52) ( 77.58, 96.93)
       ( 78.88, 97.28) ( 80.20, 97.58) ( 81.52, 97.81) ( 82.85, 97.97)
       ( 84.18, 98.05) ( 85.52, 98.05) ( 86.86, 97.95) ( 88.18, 97.75)
       ( 89.2,97.6)}
\Line( 88.52, 93.84)( 88.50, 95.17) \Line( 88.50, 95.17)( 88.86, 96.50)
\Line( 88.86, 96.50)( 89.2,97.6) \Line( 88.52, 93.84)( 88.66, 92.51)
\end{picture}  
\begin{picture}(120.0,120.0)(-35,60)
\LinAxis(  0.0,  0.0)(120.0,  0.0)(1.0,1,3,0,1)
\LinAxis(  0.0,  0.0)(  0.0,120.0)(1.0,1,-3,0,1)
\LinAxis(  0.0,120.0)(120.0,120.0)(1.0,1,-3,0,1)
\LinAxis(120.0,  0.0)(120.0,120.0)(1.0,1,3,0,1)
\Text(  0.0,-10)[]{$ -4$} \Text( 53.3,-10)[]{$  0$} \Text(120.0,-10)[]{$  5$}
\Text(-10,  0.0)[]{$  7$} \Text(-10,120.0)[]{$ 16$}
\Text(60,-30)[]{e: sheet 3, double loop}
\DashLine( 53.3,0)( 53.3,120){1.5}
\Curve{( 80.53, 10.09) ( 80.58,  8.75) ( 80.79,  7.43) ( 81.18,  6.15)}
\Curve{( 80.53, 10.09) ( 80.61, 11.43) ( 80.82, 12.76) ( 81.16, 14.05)
       ( 81.62, 15.32) (82.43,16.2)}
\Curve{( 52.70, 42.16) ( 52.74, 41.74) ( 53.26, 40.45) ( 53.84, 39.14)
       ( 54.43, 37.91) ( 55.09, 36.67) ( 55.76, 35.50) ( 56.49, 34.33)
       ( 57.27, 33.17) ( 58.07, 32.09) ( 58.91, 31.03) ( 59.79, 30.00)
       ( 60.72, 29.00) ( 61.68, 28.04) ( 62.67, 27.12) ( 63.69, 26.25)
       ( 64.76, 25.41) ( 65.87, 24.59) ( 66.98, 23.83) ( 68.11, 23.11)
       ( 69.29, 22.42) ( 70.47, 21.76) ( 71.66, 21.14) ( 72.86, 20.55)
       ( 74.07, 19.98) ( 75.29, 19.42) ( 76.52, 18.87) ( 77.75, 18.33)
       ( 78.98, 17.78) ( 80.21, 17.22) ( 81.43, 16.64) (82.43,16.2)}
\Curve{( 52.70, 42.16) ( 53.98, 42.67) ( 55.29, 43.24) ( 56.53, 43.85)
       ( 57.71, 44.47) ( 58.91, 45.17) ( 60.04, 45.89) ( 61.18, 46.67)
       ( 62.26, 47.47) ( 63.34, 48.34) ( 64.36, 49.21) ( 65.37, 50.15)
       ( 66.31, 51.11) ( 67.24, 52.12) ( 68.11, 53.14) ( 68.96, 54.23)
       ( 69.74, 55.32) ( 70.50, 56.47) ( 71.19, 57.62) ( 71.85, 58.83)
       ( 72.44, 60.04) ( 72.99, 61.30) ( 73.48, 62.57) ( 73.91, 63.88)
       ( 74.29, 65.18) ( 74.59, 66.48) ( 74.84, 67.83) ( 75.03, 69.16)
       ( 75.16, 70.54) ( 75.23, 71.90) ( 75.24, 73.24)}
\Curve{(  6.82, 76.65) (  6.88, 78.20) (  7.00, 79.74) (  7.19, 81.28)
       (  7.40, 82.60) (  7.96, 85.21) (  8.72, 87.79) (  9.67, 90.30) 
       ( 10.80, 92.74) ( 12.11, 95.08) ( 13.58, 97.31) ( 15.51, 99.75) 
       ( 17.62,101.99) ( 19.91,104.00) ( 22.35,105.78) ( 24.89,107.29) 
       ( 27.51,108.54) ( 30.16,109.51) ( 32.81,110.21) ( 35.61,110.68) 
       ( 38.49,110.88) ( 41.23,110.81) ( 44.12,110.46) ( 46.93,109.86) 
       ( 49.60,109.02) ( 52.23,107.94) ( 54.66,106.71) ( 56.99,105.30) 
       ( 59.27,103.68) ( 61.38,101.94) ( 63.37,100.04) ( 65.24, 98.02) 
       ( 66.95, 95.89) ( 68.54, 93.61) ( 69.94, 91.31) ( 71.22, 88.88) 
       ( 72.33, 86.41) ( 73.29, 83.87) ( 74.08, 81.28) ( 74.40, 79.98) 
       ( 74.67, 78.66) ( 74.90, 77.33) ( 75.07, 75.98) ( 75.18, 74.62) 
       ( 75.24, 73.24)}
\Curve{(  6.82, 76.65) (  6.83, 75.12) (  6.90, 73.59) (  7.03, 72.07)
       (  7.22, 70.56) (  7.79, 67.60) (  8.58, 64.73) (  9.58, 61.96) 
       ( 10.78, 59.31) ( 12.17, 56.80) ( 13.72, 54.45) ( 15.42, 52.26) 
       ( 17.24, 50.24) ( 19.33, 48.27) ( 21.66, 46.41) ( 24.08, 44.80) 
       ( 26.56, 43.43) ( 29.07, 42.31) ( 31.74, 41.38) ( 34.53, 40.68) 
       ( 37.26, 40.24) ( 40.03, 40.03) ( 42.81, 40.07) ( 45.56, 40.34) 
       ( 48.24, 40.83) ( 49.62, 41.17) ( 50.93, 41.56) ( 52.30, 42.01)
       ( 52.70, 42.16)}
\end{picture}
\vspace{100pt}
\caption{The numbering $(k,p)$ of the singularities, and loops around 
$x_{2,1}$ under option I.}
\label{figI}
\end{center}
\end{figure}
In \fig{figI}b we show the contour used in examining singularity $x_{2,1}$
under option I. The contour starts 
on sheet number 1 close to the origin (so that $y$ is
close to $Y_1$), moves upwards and then
to the left, circles the singularity once anti-clockwise, and returns to
its starting point by the same route in order to enable us to determine
the resulting Riemann sheet number. \fig{figI}c shows the corresponding path
in the $y$ plane. It ends again close to $Y_1$ so that, {\em for this
choice of contour and its induced branch structure\/} (indicated
in the figure), sheet 1 does not have a branch point at $x_{2,1}$.
\fig{figI}d shows what happens if, instead of sheet number 1, we start at
sheet number 3: the $y$ track starts then at close to $Y_3$, but ends up
close to $Y_5$, so that we conclude that sheets 3 and 5 are connected
at $x_{2,1}$. If we run through the whole contour twice, we get the $y$
track presented in \fig{figI}e, where the $y$ track ends up again at $Y_3$
as expected for a square root branch cut.
\begin{figure}  
\begin{center}  
\begin{picture}(120.0,120.0)(25,0)
\LinAxis(  0.0,  0.0)(120.0,  0.0)(1.0,1,3,0,1)
\LinAxis(  0.0,  0.0)(  0.0,120.0)(1.0,1,-3,0,1)
\LinAxis(  0.0,120.0)(120.0,120.0)(1.0,1,-3,0,1)
\LinAxis(120.0,  0.0)(120.0,120.0)(1.0,1,3,0,1)
\Text(  0.0,-10)[]{$-5$} \Text(100.0,-10)[]{$0$} \Text(120.0,-10)[]{$1$}
\Text(-10,  0.0)[]{$-1$} \Text(-10, 20.0)[]{$0$} \Text(-10,120.0)[]{$5$}
\Text(60,-30)[]{a: $x$-plane}
\DashLine(100,0)(100,120){1.5} \DashLine(0,20)(120,20){1.5}
\Line( 99.00, 21.00)( 29.10, 21.00) \Line( 29.10, 21.00)( 29.10, 80.90)
\CArc( 29.10, 90.90)(10,0,360)
\Vertex( 49.87, 70.13){1}\DashLine( 49.87, 70.13)( 49.87,120){6}
\Vertex( 29.10, 90.90){1}\DashLine( 29.10, 90.90)( 29.10,120){6}
\Vertex( 13.17,106.83){1}\DashLine( 13.17,106.83)( 13.17,120){6}
\end{picture} 
\begin{picture}(120.0,120.0)(-5,0)
\LinAxis(  0.0,  0.0)(120.0,  0.0)(1,1,3,0,1)
\LinAxis(  0.0,  0.0)(  0.0,120.0)(1,1,-3,0,1)
\LinAxis(  0.0,120.0)(120.0,120.0)(1,1,-3,0,1)
\LinAxis(120.0,  0.0)(120.0,120.0)(1,1,3,0,1)
\Text(  0.0,-10)[]{$-10$} \Text( 75.0,-10)[]{$  0$} \Text(120.0,-10)[]{$  6$}
\Text(-10,  7.5)[]{$  0$} \Text(-10,120.0)[]{$ 15$}
\Text(60,-30)[]{b: sheet $1$, single loop}       
\DashLine(75.0,0)(75.0,120){1.5} \DashLine(0,7.5)(120,7.5){1.5}
\Line(20.39,  8.99)(74.63,  7.88)       
\Curve{( 20.39,  8.99) ( 20.43, 10.58) ( 20.50, 12.18) ( 20.59, 13.77) 
       ( 20.71, 15.36) ( 20.85, 16.96) ( 21.02, 18.55) ( 21.22, 20.14) 
       ( 21.45, 21.73) ( 21.70, 23.33) ( 21.98, 24.92) ( 22.29, 26.51) 
       ( 22.62, 28.10) ( 22.98, 29.69) ( 23.37, 31.28) ( 23.78, 32.87) 
       ( 24.23, 34.46) ( 24.70, 36.06) ( 25.17, 37.57) ( 25.64, 39.00) 
       ( 26.13, 40.43) ( 26.64, 41.86) ( 27.18, 43.29) ( 27.74, 44.73) 
       ( 28.32, 46.16) ( 28.92, 47.59) ( 29.54, 49.03) ( 30.19, 50.46) 
       ( 30.86, 51.90) ( 31.55, 53.34) ( 32.26, 54.77) ( 32.99, 56.21) 
       ( 33.74, 57.65) ( 34.51, 59.09) ( 35.31, 60.53) ( 36.12, 61.97) 
       ( 36.96, 63.42) ( 37.81, 64.86) ( 38.68, 66.30) ( 39.47, 67.58) 
       ( 40.28, 68.86) ( 41.10, 70.13) ( 41.94, 71.41) ( 42.79, 72.68) 
       ( 43.66, 73.94) ( 44.54, 75.21) ( 45.44, 76.46) ( 46.36, 77.72) 
       ( 47.30, 78.96) ( 48.25, 80.20) ( 49.23, 81.43) ( 50.23, 82.65) 
       ( 51.25, 83.86) ( 52.31, 85.06) ( 53.39, 86.25) ( 54.51, 87.44) 
       ( 55.08, 87.88) ( 55.63, 87.36) ( 56.27, 86.81) ( 56.93, 86.29) 
       ( 57.60, 85.81) ( 58.29, 85.36) ( 59.68, 84.56) ( 61.09, 83.91) 
       ( 62.59, 83.36) ( 64.16, 82.93) ( 65.70, 82.66) ( 67.28, 82.53) 
       ( 68.85, 82.53) ( 70.41, 82.66) ( 71.94, 82.93) ( 73.47, 83.32)
       ( 74.93, 83.82) ( 76.35, 84.44) ( 77.74, 85.17) ( 79.06, 86.00)
       ( 80.31, 86.93) ( 81.49, 87.95) ( 82.58, 89.05) ( 83.58, 90.23)
       ( 84.48, 91.49) ( 85.27, 92.81) ( 85.94, 94.20) ( 86.48, 95.62)
       ( 86.89, 97.09) ( 87.17, 98.60) ( 87.25, 99.35) ( 87.30,100.12) 
       ( 87.32,100.88)}
\Curve{( 86.75,105.39) ( 86.92,104.66) ( 87.06,103.91) ( 87.18,103.16)
       ( 87.26,102.40) ( 87.31,101.63) ( 87.32,100.88)}
\Curve{( 86.75,105.39) 
       ( 87.51,105.64) ( 88.25,105.83)
       ( 88.98,106.02) ( 89.71,106.20) ( 90.45,106.40) ( 91.17,106.60)
       ( 91.89,106.82) ( 92.62,107.05) ( 93.33,107.30) ( 94.03,107.57)
       ( 94.72,107.88) ( 95.39,108.22) ( 96.04,108.61)} 
\Curve{( 94.99,111.42) ( 95.16,110.69) ( 95.41,109.98) ( 95.74,109.30) 
       ( 96.04,108.61)}
\end{picture} 
\begin{picture}(120.0,120.0)(-35,0)
\LinAxis(  0.0,  0.0)(120.0,  0.0)(1.0,1,3,0,1)
\LinAxis(  0.0,  0.0)(  0.0,120.0)(1.0,1,-3,0,1)
\LinAxis(  0.0,120.0)(120.0,120.0)(1.0,1,-3,0,1)
\LinAxis(120.0,  0.0)(120.0,120.0)(1.0,1,3,0,1)
\Text(  0.0,-10)[]{$  0.5$} \Text(120.0,-10)[]{$  3.5$}
\Text(-12,  2.0)[]{$  4.5$} \Text(-12,120.0)[]{$  7.5$}
\Text(60,-30)[]{c: sheet $3$, single loop}
\Curve{( 63.54,118.45) ( 64.53,114.57) ( 66.07,110.87) ( 68.01,107.34)
       ( 70.30,104.02) ( 72.89,100.93) ( 75.73, 98.09) ( 78.78, 95.48)}
\Curve{( 56.52, 49.31) ( 56.69, 53.36) ( 57.23, 57.39) ( 58.11, 61.31)
       ( 59.31, 65.14) ( 60.81, 68.87) ( 62.57, 72.51) ( 64.55, 76.01)
       ( 66.71, 79.41) ( 69.01, 82.72) ( 71.39, 85.94) ( 73.84, 89.10)
       ( 76.34, 92.26) ( 78.78, 95.48)}
\Curve{( 56.52, 49.31) ( 56.71, 45.31) ( 57.26, 41.27) ( 58.13, 37.34)
       ( 58.18, 36.81)}
\Curve{( 48.71, 24.70) ( 48.92, 26.75) ( 49.46, 28.73) ( 50.30, 30.58)
       ( 51.46, 32.33) ( 52.85, 33.84) ( 54.50, 35.12) ( 56.27, 36.10)
       ( 58.18, 36.81)}
\Curve{( 48.71, 24.70) ( 48.83, 22.66) ( 49.26, 20.68) ( 50.03, 18.73)
       ( 51.06, 16.97) ( 52.39, 15.36) ( 53.90, 14.01) ( 55.63, 12.89)
       ( 57.52, 12.04) ( 59.45, 11.51) ( 61.45, 11.27) ( 63.48, 11.33)
       ( 65.48, 11.71) ( 67.41, 12.39) ( 69.23, 13.36) ( 70.82, 14.57)
       ( 72.23, 16.03) ( 73.41, 17.71) ( 74.29, 19.52) ( 74.89, 21.50)
       ( 75.16, 23.53)}
\Curve{( 58.18, 36.81) ( 58.97, 37.01) ( 60.96, 37.29) ( 62.98, 37.28)
       ( 64.98, 36.96) ( 66.91, 36.34) ( 68.73, 35.45) ( 70.39, 34.28)
       ( 71.87, 32.86) ( 73.09, 31.27) ( 74.06, 29.48) ( 74.73, 27.59)
       ( 75.16, 23.53)}
\end{picture}
\vspace{40pt}
\caption{Loops around $x_{2,1}$ under option II.}
\label{figII}
\end{center}
\end{figure}
Under option II, we rather use the contour indicated in \fig{figII}a, which
first moves to the left and then upwards. \fig{figII}b shows the resulting
$y$ path, which does not return to $Y_1$ but rather to $Y_5$, indicating
that under this choice of contour the sheets labeled 1 and 5 are connected
at $x_{2,1}$. \fig{figII}c shows that, now, sheet 3 is insensitive to
this singularity.

In this way we have mapped the various singularities around the origin.
\begin{table}
\begin{center}
\begin{tabular}{|l|c|c|c|c|c|c|c|c|}
\hline 
$k$   & $x_{k,1}$ & $\ep_{k,1}$ & $x_{k,2}$ & $\ep_{k,2}$ &
$x_{-k,1}$ & $\ep_{-k,1}$ & $x_{-k,2}$ & $\ep_{-k,2}$ \\
\hline
1     & (1,3)  & -1  & (-1,3)  & -1  & (-1,-3) & -1  & (1,-3)  & -1  \\
2     & (3,5)  & -1  & (3,5)   & -1  & (-3,-5) & -1  & (-3,-5) & -1  \\
3     & (5,7)  & -1  & (5,7)   & -1  & (-5,-7) & -1  & (-5,-7) & -1  \\
4     & (7,9)  & -1  & (7,9)   & -1  & (-7,-9) & -1  & (-7,-9) & -1  \\
5     & (9,11) & -1  & (9,11)  & -1  & (-9,-11)& -1  & (-9,-11)& -1  \\
\hline 
\end{tabular}
\caption[.]{Sheets connected at the first few singularities (option I), and
the corresponding value for $\epsilon$.}
\label{table2}
\end{center}
\end{table}
In \tab{table2} we present the pairs of sheets that are pairwise connected
at the first few singularities, under option I, and the observed value for
$\ep$, which turns out to be $-1$ in all cases.
We point out that at each singularity
only two sheets out of all infinitely many are connected. Note the
somewhat atypical situation at the lowest-lying singularities $x_{1,\pm1}$
and $x_{-1,\pm1}$.
\begin{table}
\begin{center}
\begin{tabular}{|l|c|c|c|c|c|c|c|c|}
\hline 
$k$   & $x_{k,1}$ & $\ep_{k,1}$ & $x_{k,2}$ & $\ep_{k,2}$ &
$x_{-k,1}$ & $\ep_{-k,1}$ & $x_{-k,2}$ & $\ep_{-k,2}$ \\
\hline
1     & (1,3) & -1   & (-1,3) & -1  & (-1,-3) & -1  & (1,-3) & -1   \\
2     & (1,5) & -1   & (-1,5) & -1  & (-1,-5) & -1  & (1,-5) & -1   \\
3     & (1,7) & -1   & (-1,7) & -1  & (-1,-7) & -1  & (1,-7) & -1   \\
4     & (1,9) & -1   & (-1,9) & -1  & (-1,-9) & -1  & (1,-9) & -1   \\
5     & (1,11)& -1   & (-1,11)& -1  & (-1,-11)& -1  & (1,-11)& -1   \\
\hline
\end{tabular}
\caption[.]{Sheets connected at the first few singularities (option II), and
the corresponding value for $\epsilon$.}
\label{table3}
\end{center}
\end{table}
The alternative option II results in \tab{table3}. Note that the higher-lying
singularities now show a sheet structure similar to the lowest ones.
In fact, this is the choice that corresponds most directly to the
analysis of the sheet structure of the bootstrap equation in
\cite{hagedorn}, with of course the extra complication in the fact
that the bootstrap equation (\ref{bootstrap}) has $m=1$ while for $F_1$,
$m=2$. Note that, once again, $\ep=-1$ in all cases.

\subsection{Asymptotic behavior of the series expansion coefficients}
We shall now illustrate how the information on the $x_{k,p}$ and $\beta_{k,p}$
allows us to compute the asymptotic behavior of the series expansion
coefficients $\alpha_n$.

\paragraph{First Riemann sheet.} In this sheet, the singularities closest
to the origin, and their corresponding $\beta$'s are
\begin{align}
&x_{1,1}  =  \sqrt{4\pi}\exp(3i\pi/4)\;\;\;,\;\;\;
\beta_{1,1} = \sqrt{8\pi}\exp(-3i\pi/4)\;\;,\nl
&x_{-1,2}  =  \sqrt{4\pi}\exp(-3i\pi/4)\;\;\;,\;\;\;
\beta_{-1,2} = \sqrt{8\pi}\exp(3i\pi/4)\;\;.
\end{align}
Using \eqn{asymptot}, we see that the asymptotic form of the
coefficients on sheet 1 is given by
\begin{align}
&\alpha_n^{(1)}  \sim  \alasym_n\;\;,\nl
&\alasym_n  =  \frac{2}{ n^{3/2}(4\pi)^{n/2}}\,c_n\;\;,\nl
&c_n  =  -\sqrt{2}\cos\left(\frac{3n\pi}{4}+\frac{3\pi}{4}\right)
     =  \left\{\mbox{\begin{tabular}{ll}
 $(-)^{p}$           & $n=4p$ \\
 $0$                 & $n=4p+1$ \\
 $(-)^{p+1}$         & $n=4p+2$ \\
 $(-)^{p}\sqrt{2}$   & $n=4p+3$ \end{tabular}}\right.\;\;,
\end{align}
with integer $p$.
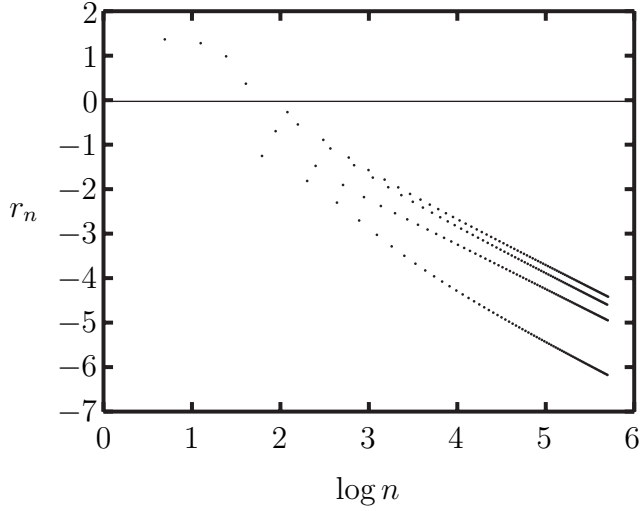
\begin{figure}
\begin{center}
\begin{picture}(200.0,151.0)(0,0)
\LinAxis(  0.0,  0.0)(200.0,  0.0)(6.0,1,3,0,1.5)
\LinAxis(  0.0,  0.0)(  0.0,151.0)(9.0,1,-3,0,1.5)
\LinAxis(  0.0,151.0)(200.0,151.0)(6.0,1,-3,0,1.5)
\LinAxis(200.0,  0.0)(200.0,151.0)(9.0,1,3,0,1.5)
\Text(  0.0,-10)[]{$0$}
\Text( 33.3,-10)[]{$1$}
\Text( 66.7,-10)[]{$2$}
\Text(100.0,-10)[]{$3$}
\Text(133.3,-10)[]{$4$}
\Text(166.7,-10)[]{$5$}
\Text(200.0,-10)[]{$6$}
\Text(-10,  0.0)[]{$-7$}
\Text(-10, 16.8)[]{$-6$}
\Text(-10, 33.6)[]{$-5$}
\Text(-10, 50.3)[]{$-4$}
\Text(-10, 67.1)[]{$-3$}
\Text(-10, 83.9)[]{$-2$}
\Text(-10,100.7)[]{$-1$}
\Text(-5,117.4)[]{$0$}
\Text(-5,134.2)[]{$1$}
\Text(-5,151.0)[]{$2$}
\Text(100,-30)[]{$\log n$}
\Text(-30,75.5)[]{$r_n$}
\Line(0,117)(200,117)
\Vertex(0.00,138.68){0.5} \Vertex(23.10,140.38){0.5}
\Vertex(36.62,138.94){0.5} \Vertex(46.21,133.98){0.5}
\Vertex(53.65,123.66){0.5} \Vertex(59.73, 96.41){0.5}
\Vertex(64.86,105.76){0.5} \Vertex(69.31,112.94){0.5}
\Vertex(73.24,108.29){0.5} \Vertex(76.75, 86.98){0.5}
\Vertex(79.93, 92.66){0.5} \Vertex(82.83,102.48){0.5}
\Vertex(85.50, 99.23){0.5} \Vertex(87.97, 78.81){0.5}
\Vertex(90.27, 85.47){0.5} \Vertex(92.42, 95.81){0.5}
\Vertex(94.44, 92.92){0.5} \Vertex(96.35, 72.07){0.5}
\Vertex(98.15, 80.89){0.5} \Vertex(99.86, 91.09){0.5}
\Vertex(101.48, 88.25){0.5} \Vertex(103.03, 66.66){0.5}
\Vertex(104.52, 77.53){0.5} \Vertex(105.93, 87.49){0.5}
\Vertex(107.30, 84.61){0.5} \Vertex(108.60, 62.31){0.5}
\Vertex(109.86, 74.81){0.5} \Vertex(111.07, 84.56){0.5}
\Vertex(112.24, 81.63){0.5} \Vertex(113.37, 58.76){0.5}
\Vertex(114.47, 72.49){0.5} \Vertex(115.52, 82.09){0.5}
\Vertex(116.55, 79.12){0.5} \Vertex(117.55, 55.78){0.5}
\Vertex(118.51, 70.46){0.5} \Vertex(119.45, 79.95){0.5}
\Vertex(120.36, 76.94){0.5} \Vertex(121.25, 53.23){0.5}
\Vertex(122.12, 68.65){0.5} \Vertex(122.96, 78.06){0.5}
\Vertex(123.79, 75.02){0.5} \Vertex(124.59, 51.01){0.5}
\Vertex(125.37, 67.02){0.5} \Vertex(126.14, 76.36){0.5}
\Vertex(126.89, 73.29){0.5} \Vertex(127.62, 49.04){0.5}
\Vertex(128.34, 65.53){0.5} \Vertex(129.04, 74.81){0.5}
\Vertex(129.73, 71.73){0.5} \Vertex(130.40, 47.27){0.5}
\Vertex(131.06, 64.16){0.5} \Vertex(131.71, 73.40){0.5}
\Vertex(132.34, 70.31){0.5} \Vertex(132.97, 45.66){0.5}
\Vertex(133.58, 62.89){0.5} \Vertex(134.18, 72.10){0.5}
\Vertex(134.77, 68.99){0.5} \Vertex(135.35, 44.19){0.5}
\Vertex(135.92, 61.71){0.5} \Vertex(136.48, 70.89){0.5}
\Vertex(137.03, 67.77){0.5} \Vertex(137.57, 42.84){0.5}
\Vertex(138.10, 60.61){0.5} \Vertex(138.63, 69.76){0.5}
\Vertex(139.15, 66.64){0.5} \Vertex(139.65, 41.59){0.5}
\Vertex(140.16, 59.57){0.5} \Vertex(140.65, 68.71){0.5}
\Vertex(141.14, 65.58){0.5} \Vertex(141.62, 40.42){0.5}
\Vertex(142.09, 58.60){0.5} \Vertex(142.56, 67.71){0.5}
\Vertex(143.02, 64.58){0.5} \Vertex(143.47, 39.32){0.5}
\Vertex(143.92, 57.68){0.5} \Vertex(144.36, 66.78){0.5}
\Vertex(144.79, 63.63){0.5} \Vertex(145.22, 38.29){0.5}
\Vertex(145.65, 56.80){0.5} \Vertex(146.07, 65.89){0.5}
\Vertex(146.48, 62.74){0.5} \Vertex(146.89, 37.32){0.5}
\Vertex(147.29, 55.97){0.5} \Vertex(147.69, 65.05){0.5}
\Vertex(148.09, 61.89){0.5} \Vertex(148.48, 36.41){0.5}
\Vertex(148.86, 55.18){0.5} \Vertex(149.24, 64.24){0.5}
\Vertex(149.62, 61.08){0.5} \Vertex(149.99, 35.54){0.5}
\Vertex(150.36, 54.43){0.5} \Vertex(150.73, 63.48){0.5}
\Vertex(151.09, 60.31){0.5} \Vertex(151.44, 34.71){0.5}
\Vertex(151.80, 53.70){0.5} \Vertex(152.15, 62.74){0.5}
\Vertex(152.49, 59.58){0.5} \Vertex(152.83, 33.92){0.5}
\Vertex(153.17, 53.01){0.5} \Vertex(153.51, 62.04){0.5}
\Vertex(153.84, 58.87){0.5} \Vertex(154.17, 33.16){0.5}
\Vertex(154.49, 52.34){0.5} \Vertex(154.81, 61.37){0.5}
\Vertex(155.13, 58.20){0.5} \Vertex(155.45, 32.44){0.5}
\Vertex(155.76, 51.70){0.5} \Vertex(156.07, 60.72){0.5}
\Vertex(156.38, 57.54){0.5} \Vertex(156.68, 31.75){0.5}
\Vertex(156.98, 51.09){0.5} \Vertex(157.28, 60.10){0.5}
\Vertex(157.58, 56.92){0.5} \Vertex(157.87, 31.08){0.5}
\Vertex(158.16, 50.49){0.5} \Vertex(158.45, 59.50){0.5}
\Vertex(158.74, 56.31){0.5} \Vertex(159.02, 30.44){0.5}
\Vertex(159.30, 49.92){0.5} \Vertex(159.58, 58.92){0.5}
\Vertex(159.86, 55.73){0.5} \Vertex(160.13, 29.82){0.5}
\Vertex(160.41, 49.36){0.5} \Vertex(160.68, 58.35){0.5}
\Vertex(160.94, 55.17){0.5} \Vertex(161.21, 29.23){0.5}
\Vertex(161.47, 48.82){0.5} \Vertex(161.73, 57.81){0.5}
\Vertex(161.99, 54.62){0.5} \Vertex(162.25, 28.65){0.5}
\Vertex(162.51, 48.30){0.5} \Vertex(162.76, 57.29){0.5}
\Vertex(163.01, 54.10){0.5} \Vertex(163.26, 28.10){0.5}
\Vertex(163.51, 47.80){0.5} \Vertex(163.76, 56.78){0.5}
\Vertex(164.00, 53.58){0.5} \Vertex(164.24, 27.56){0.5}
\Vertex(164.48, 47.30){0.5} \Vertex(164.72, 56.28){0.5}
\Vertex(164.96, 53.09){0.5} \Vertex(165.19, 27.04){0.5}
\Vertex(165.43, 46.83){0.5} \Vertex(165.66, 55.80){0.5}
\Vertex(165.89, 52.61){0.5} \Vertex(166.12, 26.53){0.5}
\Vertex(166.35, 46.36){0.5} \Vertex(166.57, 55.33){0.5}
\Vertex(166.80, 52.14){0.5} \Vertex(167.02, 26.04){0.5}
\Vertex(167.24, 45.91){0.5} \Vertex(167.46, 54.88){0.5}
\Vertex(167.68, 51.68){0.5} \Vertex(167.90, 25.56){0.5}
\Vertex(168.11, 45.47){0.5} \Vertex(168.33, 54.44){0.5}
\Vertex(168.54, 51.24){0.5} \Vertex(168.75, 25.10){0.5}
\Vertex(168.96, 45.05){0.5} \Vertex(169.17, 54.01){0.5}
\Vertex(169.38, 50.81){0.5} \Vertex(169.59, 24.64){0.5}
\Vertex(169.79, 44.63){0.5} \Vertex(170.00, 53.59){0.5}
\Vertex(170.20, 50.38){0.5} \Vertex(170.40, 24.20){0.5}
\Vertex(170.60, 44.22){0.5} \Vertex(170.80, 53.18){0.5}
\Vertex(171.00, 49.97){0.5} \Vertex(171.19, 23.77){0.5}
\Vertex(171.39, 43.82){0.5} \Vertex(171.58, 52.77){0.5}
\Vertex(171.78, 49.57){0.5} \Vertex(171.97, 23.36){0.5}
\Vertex(172.16, 43.43){0.5} \Vertex(172.35, 52.38){0.5}
\Vertex(172.54, 49.18){0.5} \Vertex(172.73, 22.95){0.5}
\Vertex(172.91, 43.06){0.5} \Vertex(173.10, 52.00){0.5}
\Vertex(173.28, 48.80){0.5} \Vertex(173.47, 22.55){0.5}
\Vertex(173.65, 42.68){0.5} \Vertex(173.83, 51.63){0.5}
\Vertex(174.01, 48.42){0.5} \Vertex(174.19, 22.16){0.5}
\Vertex(174.37, 42.32){0.5} \Vertex(174.55, 51.26){0.5}
\Vertex(174.72, 48.06){0.5} \Vertex(174.90, 21.78){0.5}
\Vertex(175.08, 41.97){0.5} \Vertex(175.25, 50.91){0.5}
\Vertex(175.42, 47.70){0.5} \Vertex(175.60, 21.41){0.5}
\Vertex(175.77, 41.62){0.5} \Vertex(175.94, 50.55){0.5}
\Vertex(176.11, 47.35){0.5} \Vertex(176.28, 21.04){0.5}
\Vertex(176.44, 41.28){0.5} \Vertex(176.61, 50.21){0.5}
\Vertex(176.78, 47.00){0.5} \Vertex(176.94, 20.69){0.5}
\Vertex(177.11, 40.94){0.5} \Vertex(177.27, 49.88){0.5}
\Vertex(177.43, 46.67){0.5} \Vertex(177.60, 20.34){0.5}
\Vertex(177.76, 40.61){0.5} \Vertex(177.92, 49.55){0.5}
\Vertex(178.08, 46.34){0.5} \Vertex(178.24, 20.00){0.5}
\Vertex(178.40, 40.29){0.5} \Vertex(178.55, 49.22){0.5}
\Vertex(178.71, 46.01){0.5} \Vertex(178.87, 19.66){0.5}
\Vertex(179.02, 39.98){0.5} \Vertex(179.18, 48.91){0.5}
\Vertex(179.33, 45.69){0.5} \Vertex(179.48, 19.33){0.5}
\Vertex(179.64, 39.67){0.5} \Vertex(179.79, 48.59){0.5}
\Vertex(179.94, 45.38){0.5} \Vertex(180.09, 19.01){0.5}
\Vertex(180.24, 39.36){0.5} \Vertex(180.39, 48.29){0.5}
\Vertex(180.54, 45.08){0.5} \Vertex(180.68, 18.69){0.5}
\Vertex(180.83, 39.06){0.5} \Vertex(180.98, 47.99){0.5}
\Vertex(181.12, 44.77){0.5} \Vertex(181.27, 18.38){0.5}
\Vertex(181.41, 38.77){0.5} \Vertex(181.56, 47.69){0.5}
\Vertex(181.70, 44.48){0.5} \Vertex(181.84, 18.08){0.5}
\Vertex(181.99, 38.48){0.5} \Vertex(182.13, 47.40){0.5}
\Vertex(182.27, 44.19){0.5} \Vertex(182.41, 17.78){0.5}
\Vertex(182.55, 38.20){0.5} \Vertex(182.69, 47.12){0.5}
\Vertex(182.83, 43.90){0.5} \Vertex(182.96, 17.49){0.5}
\Vertex(183.10, 37.92){0.5} \Vertex(183.24, 46.84){0.5}
\Vertex(183.38, 43.62){0.5} \Vertex(183.51, 17.20){0.5}
\Vertex(183.65, 37.65){0.5} \Vertex(183.78, 46.56){0.5}
\Vertex(183.91, 43.35){0.5} \Vertex(184.05, 16.91){0.5}
\Vertex(184.18, 37.38){0.5} \Vertex(184.31, 46.29){0.5}
\Vertex(184.45, 43.08){0.5} \Vertex(184.58, 16.63){0.5}
\Vertex(184.71, 37.11){0.5} \Vertex(184.84, 46.03){0.5}
\Vertex(184.97, 42.81){0.5} \Vertex(185.10, 16.36){0.5}
\Vertex(185.23, 36.85){0.5} \Vertex(185.36, 45.76){0.5}
\Vertex(185.48, 42.55){0.5} \Vertex(185.61, 16.09){0.5}
\Vertex(185.74, 36.59){0.5} \Vertex(185.87, 45.51){0.5}
\Vertex(185.99, 42.29){0.5} \Vertex(186.12, 15.82){0.5}
\Vertex(186.24, 36.34){0.5} \Vertex(186.37, 45.25){0.5}
\Vertex(186.49, 42.03){0.5} \Vertex(186.61, 15.56){0.5}
\Vertex(186.74, 36.09){0.5} \Vertex(186.86, 45.00){0.5}
\Vertex(186.98, 41.78){0.5} \Vertex(187.10, 15.30){0.5}
\Vertex(187.23, 35.84){0.5} \Vertex(187.35, 44.75){0.5}
\Vertex(187.47, 41.53){0.5} \Vertex(187.59, 15.05){0.5}
\Vertex(187.71, 35.60){0.5} \Vertex(187.83, 44.51){0.5}
\Vertex(187.95, 41.29){0.5} \Vertex(188.06, 14.80){0.5}
\Vertex(188.18, 35.36){0.5} \Vertex(188.30, 44.27){0.5}
\Vertex(188.42, 41.05){0.5} \Vertex(188.53, 14.55){0.5}
\Vertex(188.65, 35.13){0.5} \Vertex(188.77, 44.03){0.5}
\Vertex(188.88, 40.81){0.5} \Vertex(189.00, 14.31){0.5}
\Vertex(189.11, 34.89){0.5} \Vertex(189.23, 43.80){0.5}
\Vertex(189.34, 40.58){0.5} \Vertex(189.45, 14.07){0.5}
\Vertex(189.57, 34.66){0.5} \Vertex(189.68, 43.57){0.5}
\Vertex(189.79, 40.35){0.5} \Vertex(189.90, 13.83){0.5}
\Vertex(190.01, 34.44){0.5} \Vertex(190.13, 43.34){0.5}
\end{picture}
\vspace{40pt}
\caption{$r_n$, defined in \eqn{defrn}, as function of $\log n$.}
\label{figrn}
\end{center}
\end{figure}
In \fig{figrn} we have plotted the observed behavior of 
\begin{equation}
r_n = \log\left(\frac{(4\pi)^{n/2}n^{3/2}}{2}
|\alpha_n-\alasym_n|\right)
\label{defrn}
\end{equation}
on the first Riemann sheet, against $\log n$. The coefficients clearly
converge to the computed behavior, and we can even distinguish that the
leading corrections go as $n^{-5/2}$; the four separate lines that emerge
are just the four different forms of $c_n$. The series expansion for Riemann
sheet $-1$ are simply obtained from
\begin{equation}
\alpha_n^{(-1)} = (-)^n\alpha_n^{(1)}\;\;.
\end{equation}
\paragraph{Higher Riemann sheets.} We first consider positive
sheet label $s=3,5,7,\ldots$ and put $k=(s-1)/2$.
We then have
\begin{equation}
x_{k,1} = -x_{k,2} = \sqrt{4\pi k}\exp(i\pi/4)\;\;\;,\;\;\;
\beta_{k,1} = \beta_{k,2} = (1+i)\sqrt{4\pi k}\;\;.
\end{equation}
As we have already seen $\alpha_n$ vanishes for odd $n$, and for
even $n$ we have the following asymptotic form:
\begin{equation}
   \alpha_{4p}^{(s)}  \sim  {2(1+i)\sqrt{k}\over(4\pi k)^{2p}}
   (-)^{p+1}\;\;,\quad
   \alpha_{4p+2}^{(s)}  \sim  {2(1-i)\sqrt{k}\over(4\pi k)^{2p+1}}
   (-)^{p}\;\;,
\end{equation}
for integer $p$. For negative $s$, we use \eqn{coefstar}, which also holds 
asymptotically.   

\section{Conclusion}
We have shown how numerical tracking can be helpful in the investigation of the 
Riemann sheet structure of the solution of certain algebraic complex equations.
Furthermore, we have shown how the series expansions around the origin on the 
different sheets and the asymptotic behavior of their coefficients can be 
determined. The results of the numerical analyses have been justified by the 
fact that only finite computer accuracy was required in the specific 
calculations.

\end{document}